\documentclass[aps,prb,preprint,showpacs,superscriptaddress]{revtex4}

\usepackage{graphicx}
\usepackage{epsfig}
\usepackage{color}
\begin{document}
\topmargin-1.0cm

\title {
The role of triplet excitons in enhancing polymer solar cell efficiency: a photo-induced absorption study}

\author {K. Yang}
\affiliation {Department of Physics and Astronomy, University of Missouri, Columbia, Missouri 65211}\author {U. Scherf}\affiliation {Bergische
Universit$\ddot{a}$t Wuppertal, Makromolekulare Chemie, Wuppertal, Germany}\author{S. Guha}\email[Corresponding author
E-mail:]{guhas@missouri.edu} \affiliation {Department of Physics and Astronomy, University of Missouri, Columbia, Missouri 65211}

\date{\today}

\begin{abstract}
Inclusion of heavy metal atoms in a polymer backbone allows transitions between the singlet and triplet manifolds. Interfacial dissociation of
triplet excitons constitutes a viable mechanism for enhancing photovoltaic (PV) efficiencies in polymer heterojunction-based solar cells. The PV
efficiency from polymer solar cells utilizing a ladder-type poly (para-phenylene) polymer (PhLPPP) with trace quantity of Pd atoms and a
fullerene derivative (PCBM) is much higher than its counterpart (MeLPPP) with no Pd atom. Evidence is presented for the formation of a weak
ground-state charge-transfer complex (CTC) in the blended films of the polymer and PCBM, using photo-induced absorption (PIA) spectroscopy. The
CTC state in MeLPPP:PCBM has a singlet character to it, resulting in a radiative recombination. In contrast, the CTC states in PhLPPP:PCBM are
more localized with a triplet character. An absorption peak at 1.65 eV is observed in PhLPPP:PCBM blend in the PIA, which may be converted to
weakly-bound polaron-pairs, contributing to the enhancement of PV efficiency.

\end{abstract}

 \maketitle
\section{Introduction} \label{sec:intro}

The development of high efficiency organic solar cells based on bulk heterojunctions (BHT) is rapidly increasing as a viable alternative to
inorganic solar cells. New routes entail development of low bandgap, \cite{peet,brabec06} and organometallic polymers,\cite{wong} small
molecule-based donor layers, \cite{thompson} tandem cell architectures, \cite{forrest} and controlling interfacial separations to match exciton
diffusion lengths to few tens of nanometers. \cite{kietzke} A common organic BHT system which gives power efficiencies in the 4 \% range
consists of poly(3-hexylthiophene) (P3HT) and [6,6]-phenyl C$_{61}$-butyric acid methyl ester (PCBM). The dynamics of photo-induced charge
separation in donor-acceptor complexes has been extensively studied \cite{liddell,janssen1,panda} and recently it has been realized that the
mechanism of charge separation, especially at a polymer-fullerene interface, is not simply a single-step electron transfer but may involve
several intermediate steps. Several works have indicated the presence of a charge transfer complex (CTC), \cite{goris,veldman,loi,haeldermans}
which is essentially a Coulomb bound polaron pair between the donor-acceptor blend. Recently, such CTCs have been identified to be both above
and below-gap excitations. \cite{drori,herz} Below-gap excitations may be further exploited for harnessing solar energy, thus enhancing the
overall photovoltaic (PV) efficiency of organic solar cells.

The synthesis and use of organometallic donor materials for efficient conversion of solar energy have been on the rise. Many of these materials
are based on small organic molecules with a combination of phosphorescent dye molecules with iridium or platinum complexes.
\cite{forrest,yang,Hu} The central idea behind the inclusion of heavy metal atoms in the polymer backbone for PV applications is that it allows
transitions between the singlet and triplet manifolds. A low concentration of metal atoms enables strong localized spin-orbit coupling. The
photovoltaic (PV) process in organic materials stems from the generation of singlet excitons upon photon absorption, which may further result in
a spin-flip due to spin-orbit coupling leading to the formation of small fractions of triplet excitons. The long lifetimes and diffusion lengths
of triplet excitons facilitate their diffusion process towards donor-acceptor junctions. Additionally, if the energy level of the lowest
unoccupied molecular orbitals (LUMO) of the acceptor material is well matched with the lowest triplet energy of the donor, the dissociation
process is favored, thus enhancing the PV efficiency in organic solar cells. Organometallic materials may be further useful in PV applications
since the triplet exciton density is enhanced due to a strong metal-induced spin-orbit coupling. Such an approach has been suggested to increase
the electroluminescence efficiency in organic light-emitting diodes (LEDs) by transferring the triplet excited state from the polymer to an
organometallic dye.\cite{baldo,tessler}

The polymer used in this study is a diphenyl-substituted ladder-type poly(para-phenylene) (PhLPPP), containing a trace concentration of
covalently bound Pd atoms. Trace quantities of Pd atoms in PhLPPP open up an efficient decay channel for migrating triplet excitons.
Additionally, the basic photophysics (photoluminescence and absorption) of this polymer is comparable to its counterpart with no Pd atoms,
methylated-LPPP (MeLPPP). The Pd content (as estimated by inductively coupled plasma optical emission spectroscopy) is $<$ 5 ppm for MeLPPP and
120-200 ppm for PhLPP, respectively. \cite{reufer} PhLPPP has served as an important system for elucidating fundamental properties of
electrophosphorescence and temperature-dependent triplet diffusion.\cite{lupton} Recently it was shown by Arif \emph{et al.} \cite{Arif} that
PhLPPP blended with PCBM results in a PV efficiency almost 10 times more than MeLPPP blended with PCBM, when utilized in a solar cell structure.
This was attributed to the additional contribution of triplet excitons in PhLPPP.

A nominal balance between singlet and triplet excitons is perhaps
one of the key parameters for enhancing solar cell efficiencies; a
very large fraction of triplet excitons may be detrimental as it may
provide a radiative route to recombination. Additionally, the
binding energy of triplet excitons is higher than that of the
singlet excitons; hence converting all singlets to triplets may not
be an ideal situation for enhancing solar cell efficiencies. Simply
mixing trace quantity of Pd complexes to a polymer does not
necessarily work since the metal atoms may be far away from the
polymer chain and may cluster. A side chain sequence during polymer
synthesis, as in PhLPPP, ensures the metal atoms to be covalently
bound to the polymer backbone.

To gain insight into the differences in PV mechanism between the
triplet-enhanced polymer, PhLPPP, and the non-triplet-enhanced
polymer, MeLPPP, we have studied the charge transfer process using
photo-induced absorption (PIA) spectroscopy. The CTC states are
formed in both polymer blends, although the nature of these states
are very different in the two materials. The paper is structured as
follows: Section II contains the experimental details. In Section
III, we highlight the optical properties including triplet-triplet
(TT) absorption results from pure polymer films, followed by PIA
results of polymer:PCBM films for different blend ratios in Section
IV. Section V is a discussion of our results, which correlates the
PIA results to the solar cell performance.

\section{Experimental Details}\label{sec:exptaldetails}
The active polymers for the PIA study included PhLPPP with palladium
atoms present in less than 1 in 1000 polymer repeat units, and
MeLPPP with no heavy metals. PCBM $>$99.5 \% was purchased from
Aldrich. PhLPPP:PCBM and MeLPPP:PCBM blends were prepared in
different weight ratios,  2:1 (33\% wt PCBM), 1:1 (50 \% wt PCBM)
and 1:2 (67 \% wt PCBM) by dissolving the polymers in chlorobenzene.
The blended films were drop-casted onto glass substrates for the PIA
experiments. This yielded film thicknesses that were at least five
to six times more than what was used in the solar cells.

Indium tin oxide spin-coated with poly(3,4-ethylene-dioxythiophene)/poly-styrene sulphonate (PEDOT-PSS) served as the hole collection electrode
and Ca/Al served as the electron collecting electrode in PV devices. The blended polymer:PCBM solutions were spincoated onto the PEDOT-PSS layer
forming film thicknesses of $\sim$ 100 nm. The entire device fabrication process was carried out in a N$_2$ glove box and the final structure
was encapsulated, such that the polymer layers were not exposed to any air/oxygen during the current-voltage measurements. Details on device
fabrication and characterization are found in Ref. (\onlinecite{Arif}).

Device efficiencies were measured in Air Mass (AM) 1.5 G
illumination condition with the light source operating at 100
mW/cm$^2$ (Oriel 96000, 150W solar simulator). The current-voltage
measurements were carried out by a Keithley 2400 sourcemeter. The
PIA spectra were measured using the 325 nm line of a HeCd laser as
the pump beam, modulated by a mechanical chopper, schematically
shown in Figure 1. A halogen lamp was used as the probe beam which
passed through the sample and was then dispersed by a Spectral
Products CM110 monochromator. The probe beam was detected with a
silicon detector. The small changes in transmission
($\bigtriangleup$T) were picked up by a Stanford Research SR 510
lock-in amplifier referenced to the chopper. All photomodulation
spectra are $\bigtriangleup$T divided by the transmission (T). The
samples were kept in an evacuated closed-cycle cryostat to avoid any
photodegradation.

\begin{figure}
\unitlength1cm
\begin{picture}(4.0,8)
\put(-4,-0.5){ \epsfig{file=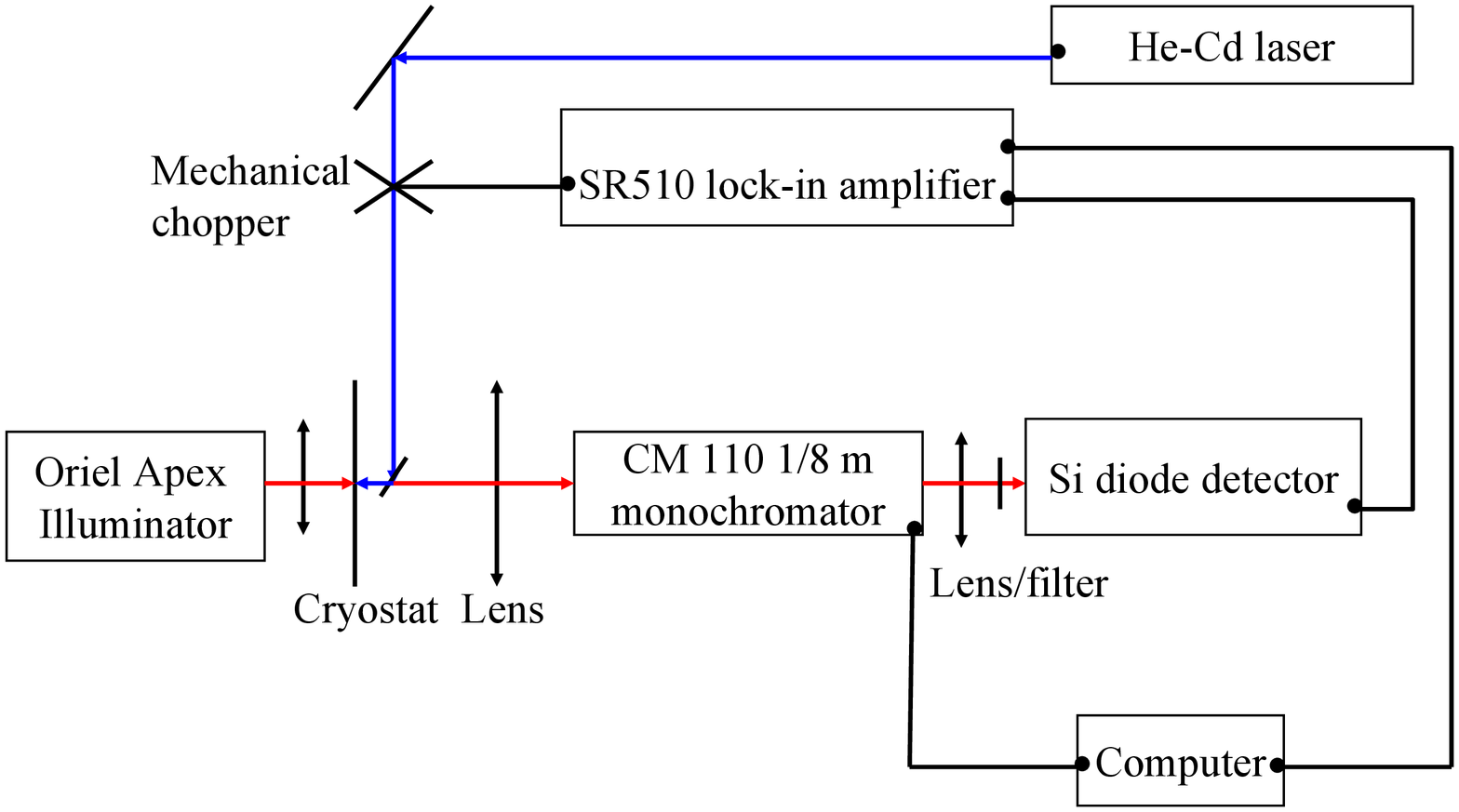, angle=0, width=12.5cm, totalheight=10.7cm}}
\end{picture}
\caption{Schematic of the experimental set up.} \label{figure1}
\end{figure}

\section{Optical Studies of PhLPPP and MeLPPP}\label{sec:PhLPPP}
\begin{figure}
\unitlength1cm
\begin{picture}(4.0,8)
\put(-4.5,-1){ \epsfig{file=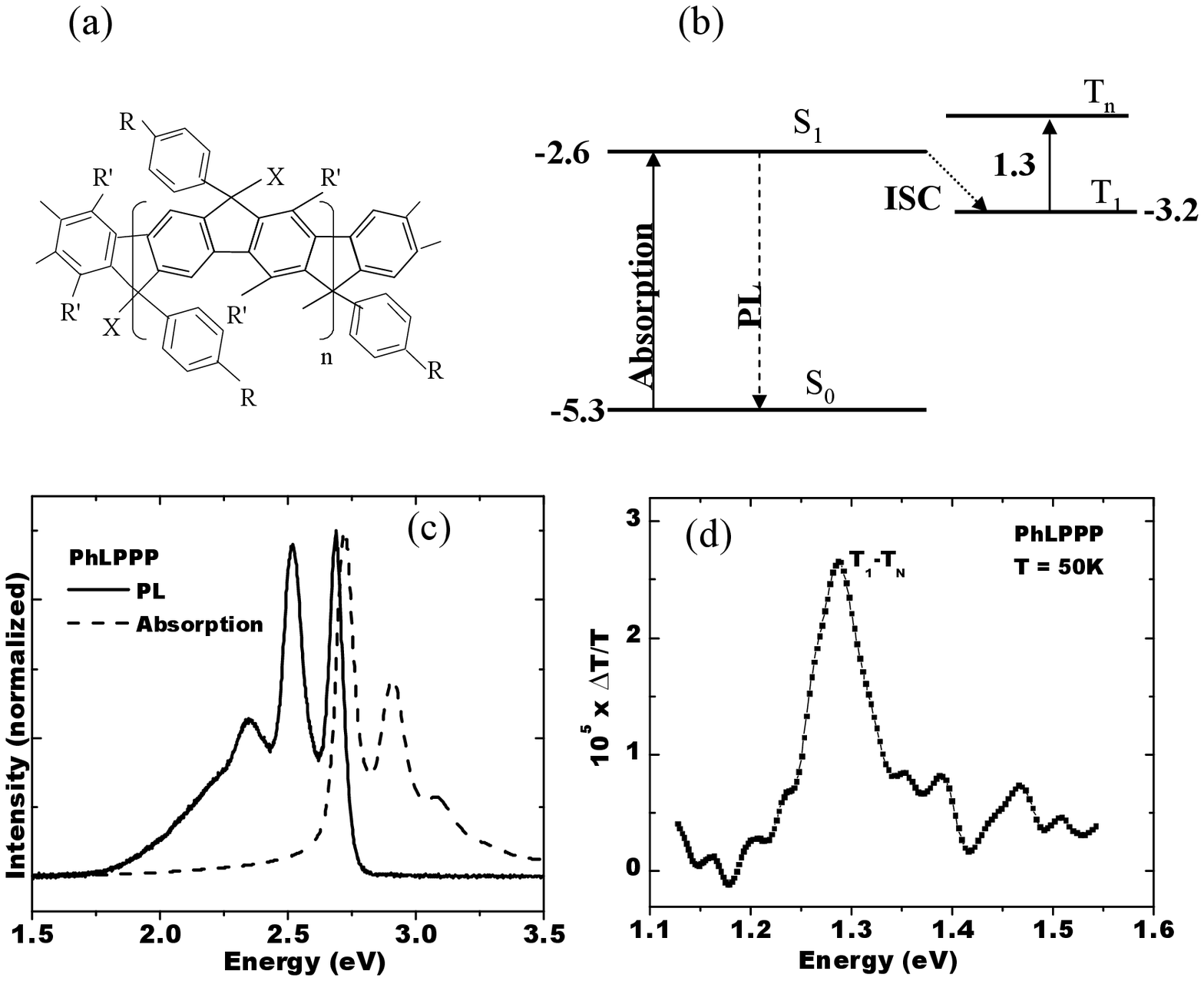, angle=0, width=14.cm, totalheight=10.7cm}}
\end{picture}
\caption{(a) Chemical structure of MeLPPP/PhLPPP; R=decyl group, R'=hexyl group,  X=Me (MeLPPP), and X=Ph (PhLPPP). (b) Energy level diagram of
PhLPPP/MeLPPP. (c) Photoluminescence and absorption spectra of a PhLPPP film. (d) Photomodulation spectrum of PhLPPP at 50 K.} \label{figure2}
\end{figure}

Figure 2 (a) and (b) schematically show the chemical structure and the energy levels of PhLPPP, respectively, which are almost identical to that
of MeLPPP. X in the chemical structure corresponds to a phenyl (Ph) group in PhLPPP and to a methyl (Me) group in MeLPPP. The highest occupied
molecular orbital (HOMO) and LUMO energy levels were previously determined using cyclic voltammetry.\cite{lupton} Neighboring phenyl rings in
MeLPPP/PhLPPP are planar due to the methyl bridges and show no torsional degree of freedom. The planarity of the phenyl rings results in high
intrachain order and consequently a low defect concentration. This is reflected in the small Stokes shift between the absorption and
photoluminescence (PL) spectra, and clear vibronic peaks in the absorption spectrum (Fig. 2(c)). We note that MeLPPP shows almost identical
PL/absorption spectra as PhLPPP. The PIA spectrum clearly shows the triplet (T$_1$)$\longrightarrow$ triplet (T$_n$) absorption peak (TT) at 1.3
eV (similar to MeLPPP) in Fig.\ref{figure2}(d). Since the intersystem cross-over (ISC) is enhanced in PhLPPP, unlike MeLPPP the TT absorption
peak is observed even at room temperature (RT). The electron-electron interactions in organic semiconductors split the lowest singlet and
triplet states by the exchange energy $\bigtriangleup$E$_{ST}$. K\"{o}hler \emph{et al}. have shown that in phenylene-based conjugated polymers,
$\bigtriangleup$E$_{ST}$ has an almost constant value of 0.7$\pm$0.1 eV, \cite{koehler} implying that the T$_1$ state in these polymers is at
-3.2 eV.

\begin{figure}
\unitlength1cm
\begin{picture}(4.0,7)
\put(-4.5,-1){ \epsfig{file=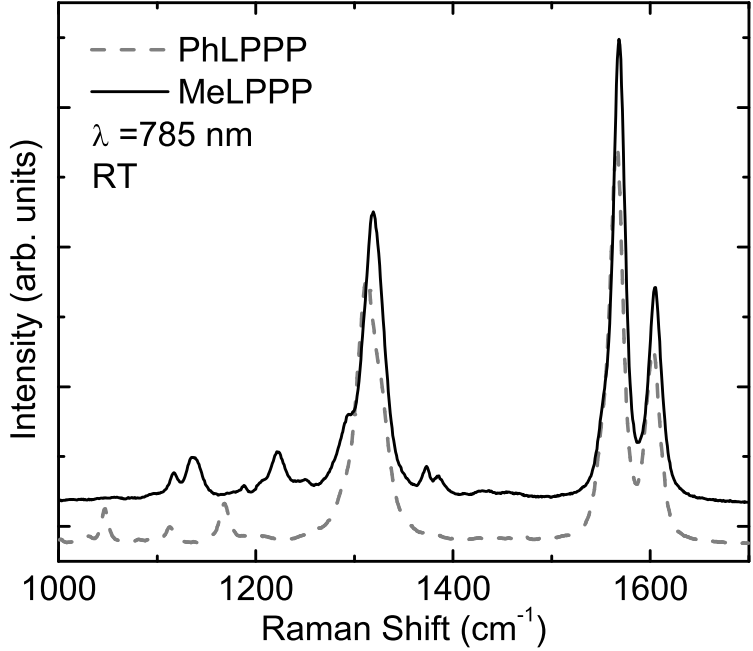, angle=0, width=12.cm, totalheight=10.7cm}}
\end{picture}
\caption{Raman spectra of PhLPPP (dotted line) and MeLPPP (bold
line) films at RT measured with an excitation wavelength of 785 nm.
The two spectra are vertically displaced for clarity.}
\label{figure3}
\end{figure}

The Raman spectra of MeLPPP and PhLPPP are similar indicating a structural similarity of both polymers. The Raman spectra of both polymers at RT
in the 1000-1700 cm$^{-1}$ range are shown in Fig.\ref{figure3}. For side-group substituted polyphenyl polymers, the Raman frequencies in the
1000-1200 cm$^{–1}$ region are most sensitive to side-group substitution. Since PhLPPP has dipenhyl substitution in its side chain (which was
necessary for the incorporation of Pd in the backbone), the two polymers show slight differences in this region of the Raman spectrum. The 1300
cm$^{-1}$ Raman peak represents the C-C stretch backbone mode. The Raman peaks in the 1600 cm$^{-1}$ region arise from the intra-ring C-C
stretch mode of the phenyl rings. In shorter oligophenyls, this mode is a double peak structure that has been suggested to originate from a
Fermi resonance between a fundamental mode and some combination band. \cite{heimel} In longer chain PPP-type polymers the doublet in the 1600
cm$^{-1}$ region is due to a lowering of symmetry. \cite{guha1} The relative ratio of the intensities of the two Raman peaks in the 1600
cm$^{-1}$ region are similar in the two polymers. From steady state optical studies we can thus conclude that the structural and optical
properties of PhLPPP and MeLPPP are similar; therefore the differences in the photovoltaic process in the two polymers arise as a result of the
triplet excitons and/or other electronic complexes formed in the blended films.

\subsection{PIA studies of pristine PhLPPP and MeLPPP}\label{sec:asis film}
The TT absorption at 1.3 eV in PhLPPP is clearly seen at RT. The PIA
spectrum was measured both as a function of temperature and chopper
frequency. Temperature-dependent PIA studies of PhLPPP show the TT
absorption to blue-shift with increasing temperatures, suggesting a
localization of the excitons on smaller chain segments with large
exchange splitting energy. This work is described in detail in Ref.
(\onlinecite{kai}). By tracking the intensity of the TT absorption
as a function of the chopper frequency, lifetime of the triplet
states can be estimated. A few selected PIA spectra for various
chopper frequencies at RT are shown in the inset of Fig. 4. We find
that $\triangle$T/T vs. frequency (\emph{f}) (Fig. 4) fits with one
time constant, $\tau$, using the relation
\begin{equation}\label{1}
\Delta T/T= C/\sqrt{1+2\pi f \tau},
\end{equation}
where \emph{C} is a constant. By fitting the PhLPPP data at 300 K
with Eq. (1), the triplet lifetime was determined as $\tau = 8.7 \pm
0.4$ ms. At 50 K the lifetime was determined to be slightly higher
at 9.5 ms. Work by Reufer \emph{et al.} on PhLPPP shows that the
migration of non-thermalized triplet excitons is driven by nearest
neighbor exchange interaction where the diffusivity is temperature
independent and has an average value of 20$\times$10$^{-6}$
cm$^2$/s. \cite{reufer} This implies that triplet exciton diffusion
lengths are in the micron range making the polymer attractive for
photovoltaic applications.

\begin{figure}
\unitlength1cm
\begin{picture}(4.0,8)
\put(-4.5,-0.5){ \epsfig{file=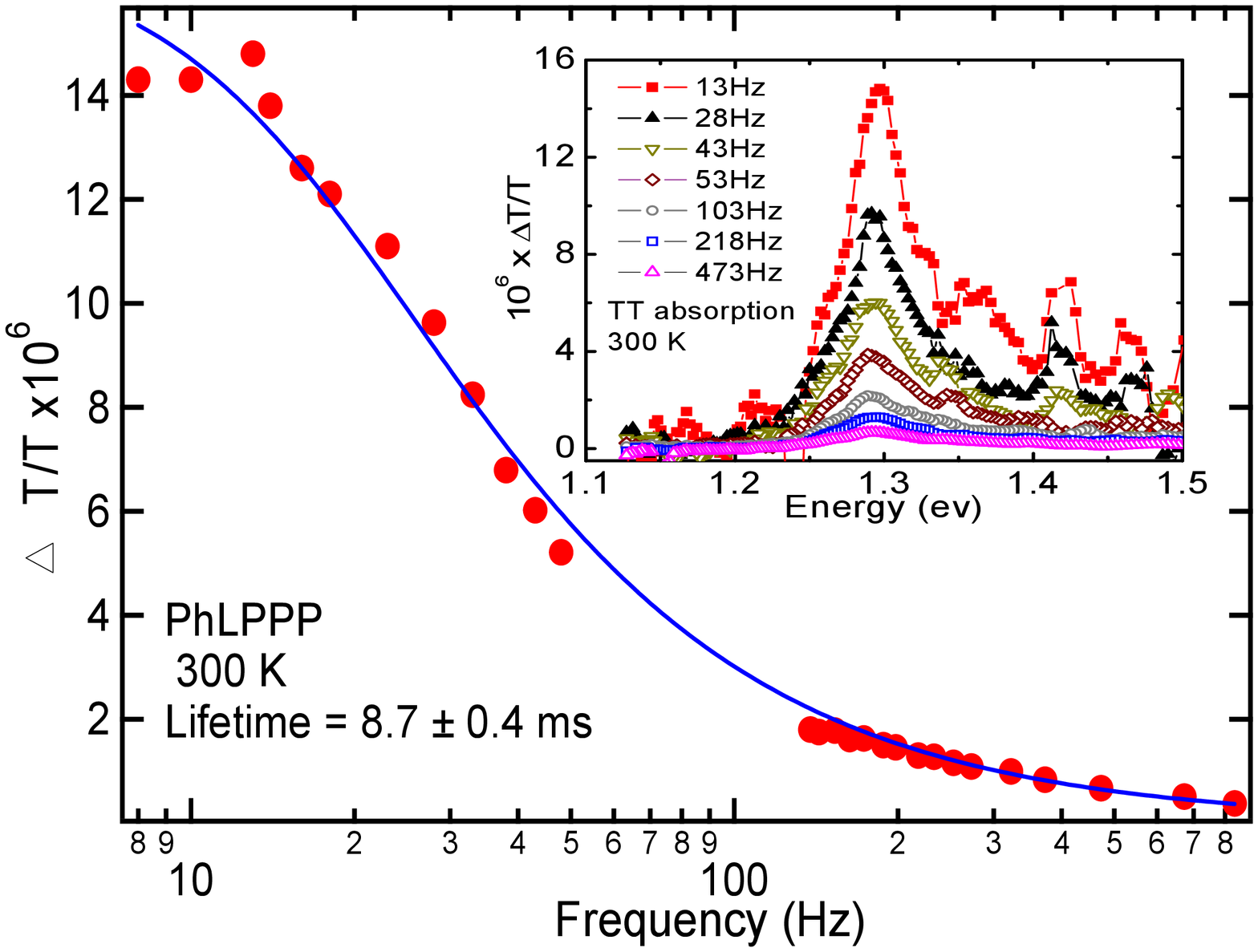, angle=0, width=12.cm, totalheight=9.7cm}}
\end{picture}
\caption{The intensity of the triplet-triplet absorption in PhLPPP
at RT as a function of the chopper frequency. The inset shows the
photommodulation spectra of PhLPPP at a few chopper frequencies.}
\label{figure4}
\end{figure}

Details of PIA spectra of pristine MeLPPP films (including photo-oxidized films and under hydrostatic pressures) were reported in Ref.
(\onlinecite{syang1}). The lifetime of triplet excitons in MeLPPP was found to be somewhat smaller ($\sim$ 4 ms).\cite{syang2} The TT absorption
in MeLPPP (shown in Fig. 5) is observed mainly at low temperatures since the ISC mechanism is much weaker in this polymer. The peak at 2.7 eV is
due to photobleaching, which results due to a depopulation of the ground state. Since both MeLPPP and PhLPPP have a very high level of purity
(with conjugation length $\sim$70), the polaronic signal is very weak in pristine polymer films. Typically, defects in the polymer backbone
resulting from an interruption of the alternating single-double bond sequence, torsion between neighboring units, or a bent arrangement of the
polymer such as an ortho- or meta- substitution on an otherwise para-substituted polymer give rise to polaron formation that can be optically
detected. Both PhLPPP and MeLPPP films were photo-oxidized by irradiating with the UV line (351.1 nm) of an Ar-ion laser in air. The resulting
polaronic absorption at 1.9 eV is shown in Fig. 5 for photo-oxidized films of PhLPPP and MeLPPP.

\begin{figure}
\unitlength1cm
\begin{picture}(4.0,7)
\put(-4.5,-1){ \epsfig{file=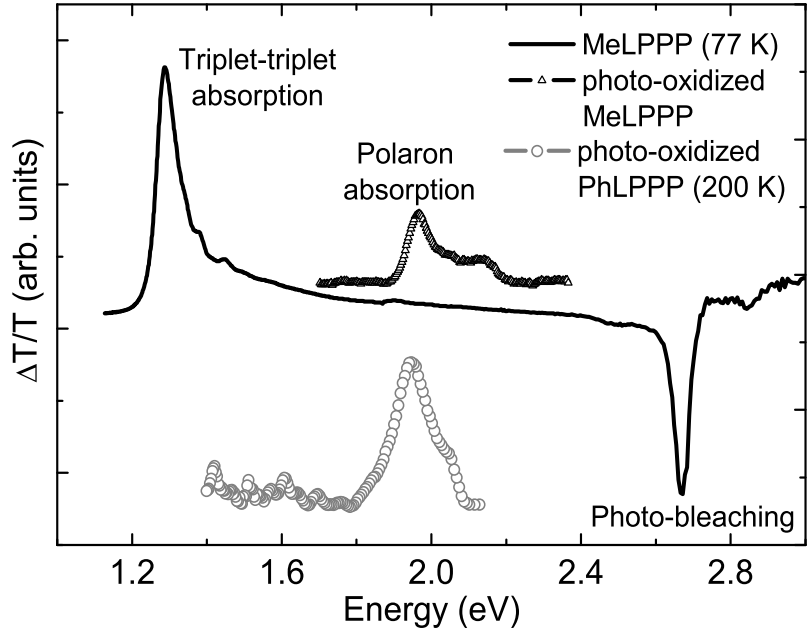, angle=0, width=12.cm, totalheight=9.7cm}}
\end{picture}
\caption{Photomodulation spectra of MeLPPP and PhLPPP. The black bold line is the PIA spectrum of MeLPPP at 77 K; the 1.3 eV and 2.65 eV
features denote the TT absorption and photobleaching, respectively. The $\triangle$ and $\circ$ symbols are the PIA spectra from photo-oxidized
MeLPPPP and PhLPPP films, respectively.} \label{figure5}
\end{figure}

\section{PIA results: polymer blends}\label{sec:PIA results}
PhLPPP:PCBM and MeLPPP:PCBM blends were prepared in different weight ratios, namely, 2:1, 1:1, and 1:2 by dissolving the polymers in
chlorobenzene. The bandgap of MeLPPP/PhLPPP is $\sim$ 2.5 eV, far from the ideal low bandgap required for high efficiency polymer-based solar
cell applications. Hence it is not surprising that power conversion efficiencies are smaller than those reported for P3HT-based solar cells.
Furthermore, for a balanced charge transport, the hole and electron mobilities should be of the same order; in PhLPPP and MeLPPP based polymers
the hole mobilities are a few orders of magnitude smaller than the electron mobilities in PCBM. Details of device properties of PhLPPP and
MeLPPP-based solar cells are given in Ref.(\onlinecite{Arif}). It should be pointed out that if charge injection/collection problems are
eliminated either by using better electrodes or by improved donor-acceptor interfaces, it is conceivable that the PV efficiency in MeLPPP-based
solar cells is considerably enhanced. Since this work compares the PIA results to actual device performance, we tabulate the power conversion
efficiencies (PCE) for the various blended solar cells in Table \ref{table1} from Ref. (\onlinecite{Arif}).

\begin {table}
\caption{Power conversion efficiency (PCE) of solar cells fabricated using different ratios of MeLPPP and PhLPPP blended with PCBM. [Ref.
(\onlinecite{Arif})]} \label{table1}
\begin{ruledtabular}
\begin{tabular}{cccc}
Device with active layer     & PCBM conc. (wt\%) &
PCE (\%)\\
\hline \
MeLPPP-PCBM(2:1)   & 33& 0.008\\
PhLPPP-PCBM (2:1)  & 33 & 0.09 \\
MeLPPP-PCBM(1:1)   & 50& 0.02\\
PhLPPP-PCBM (1:1)  & 50 & 0.23 \\
MeLPPP-PCBM(1:2)   & 67& 0.07\\
PhLPPP-PCBM (1:2)  & 67 & 0.23 \\

\end{tabular}
\end{ruledtabular}
\end{table}

Figure 6 compares the PIA spectrum in the 1.4-2.0 eV range for the 2:1 (polymer:PCBM) films. The PIA signature from the two films is vastly
different; PhLPPP blend shows a positive PIA whereas the MeLPPP sample shows a negative PIA. The positive peak at 1.65 eV in PhLPPP blend
indicates an absorption process and the negative peaks at 1.45 eV and 1.58 eV in the MeLPPP blend indicate radiative emission. Based on these
results along with the HOMO/LUMO and triplet state energy levels, we can speculate the energy levels of CTC states in both blends.

\begin{figure}
\unitlength1cm
\begin{picture}(4.0,8)
\put(-4.5,-1){ \epsfig{file=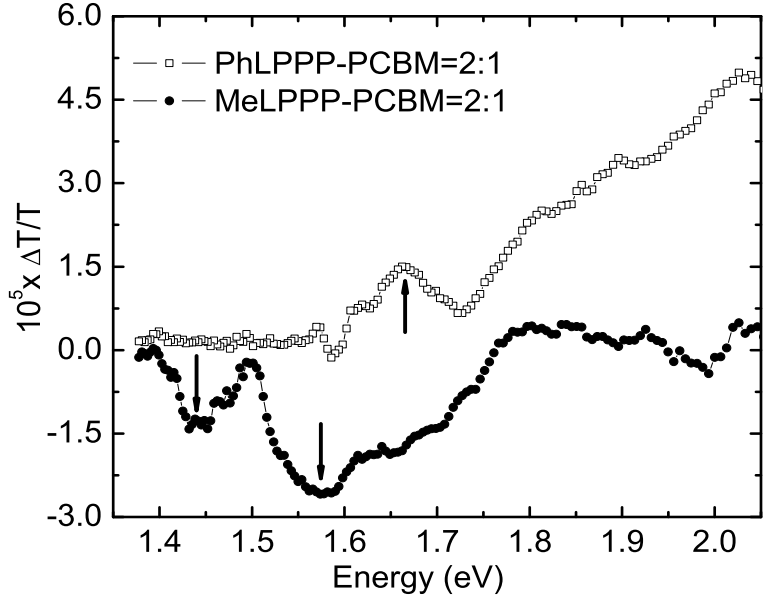, angle=0, width=12.cm, totalheight=9.7cm}}
\end{picture}
\caption{Photomodulation spectra of 2:1 PhLPPP:PCBM and MeLPPP:PCBM samples at RT. The arrows point to the main absorption and emission features
seen in both samples.} \label{figure6}
\end{figure}

\begin{figure}
\unitlength1cm
\begin{picture}(4.0,7)
\put(-4.5,-1){ \epsfig{file=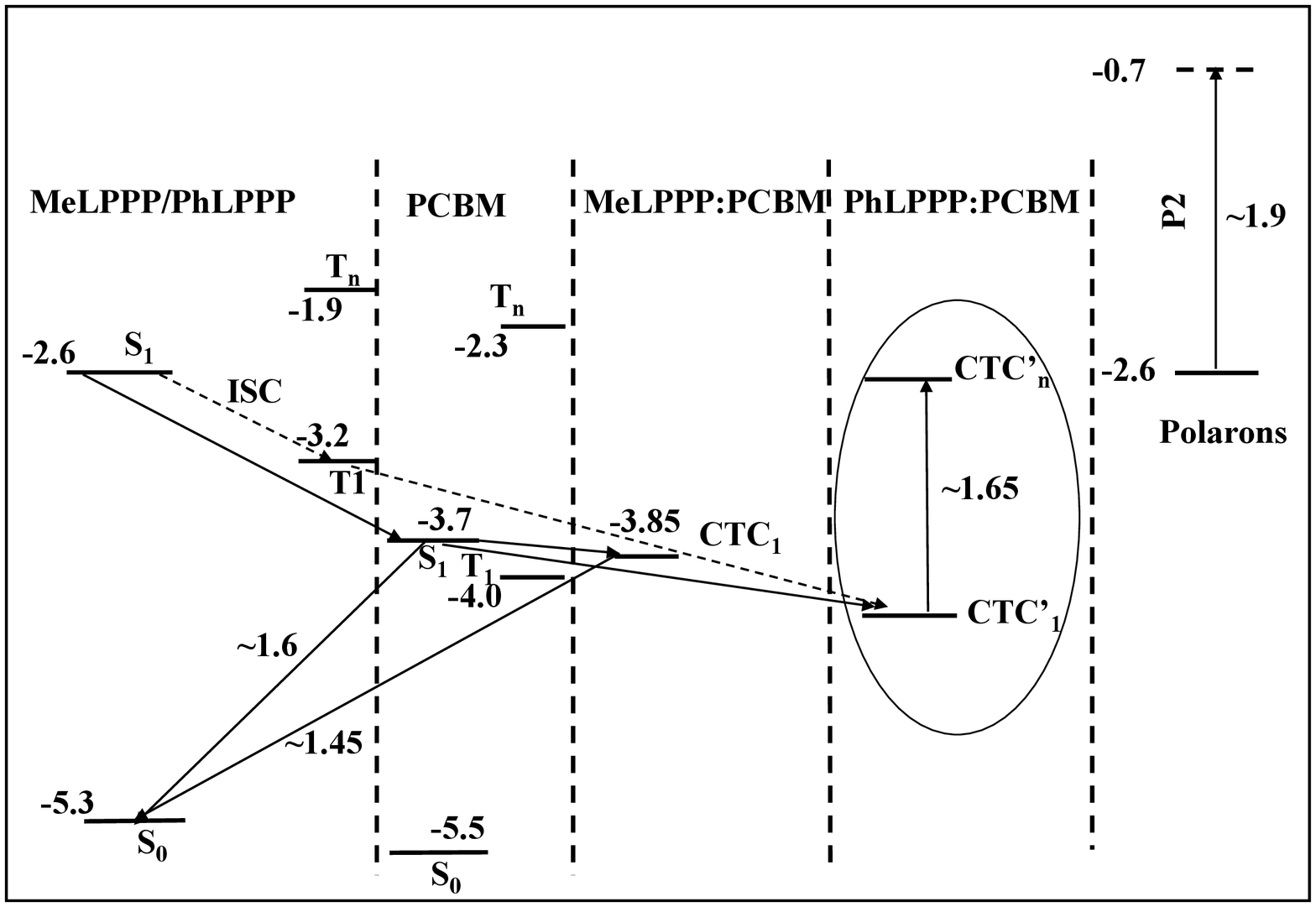, angle=0, width=12.cm,
totalheight=8.7cm}}
\end{picture}
\caption{Energy levels of the singlet and triplet states of MeLPPP/PhLPPP and PCBM. The energy levels of the charge transfer complexes are
determined from the PIA experiments.} \label{figure7}
\end{figure}

Figure \ref{figure7} schematically plots the energy levels of PhLPPP/MeLPPP and PCBM. The PCBM energy levels are taken from Refs.
(\onlinecite{smith},\onlinecite{schueppel}) The energy levels of PhLPPP and MeLPPP are similar; the main difference being a higher ISC in
PhLPPP. The transitions depicted by dotted lines involve the triplet states. In $\pi$-conjugated polymers the addition of a charge carrier onto
the backbone forms a spin 1/2 polaron with two allowed optical transitions, P$_1$ and P$_2$.\cite{fesser} Only the P$_2$ polaron absorption,
observed in photo-oxidized films (Fig. 5), is schematically shown in Fig. \ref{figure7}. As pointed out in Ref. (\onlinecite{smith}) the CTC
state may either decay radiatively or be converted into a bound polaron-pair state; the polaron-pair level must be at a slightly higher energy
since the CTC level overcomes an energy barrier for polaron-pair formation. In MeLPPP:PCBM blends the radiatve emission at 1.45 eV implies that
the CTC state is roughly at $\sim$ -3.85 eV. We note that S$_1$$\rightarrow$S$_0$ transition in PCBM is symmetry forbidden. Since the singlet
energies in PhLPPP and MeLPPP are similar, the singlet CTCs in both blends should have similar energies. The experimental results clearly point
to very different PIA features in PhLPPP:PCBM. The absorption peak at 1.65 eV may be a result of a long-lived CTC triplet state, which is
populated directly from the triplet state of PhLPPP. We denote the absorption at 1.65 eV in PhLPPP:PCBM (seen in Fig. \ref{figure6}) as a
transition between CTC$^\prime_1$ and a higher level CTC$^\prime$ state in Fig.\ref{figure7}. Also, the CTC$^\prime_1$ state most probably lies
below the S$_1$ and triplet states of the polymer and PCBM as no radiative emission is seen in the PIA spectrum. We return to this discussion in
Section \ref{sec:discussion}.

\subsection{PIA results of MeLPPP:PCBM}\label{sec:melppp-pcbm film}
\begin{figure}
\unitlength1cm
\begin{picture}(4.0,7)
\put(-4.5,-1){ \epsfig{file=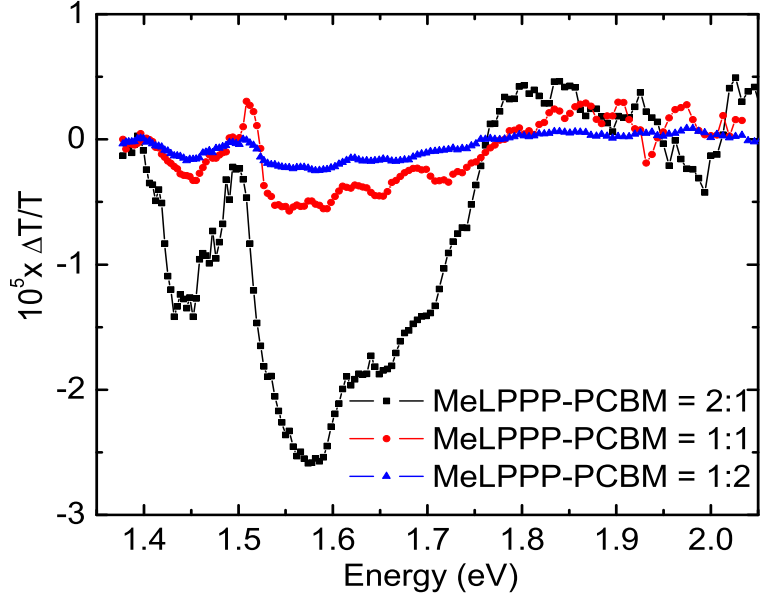, angle=0, width=12.cm, totalheight=9.7cm}}
\end{picture}
\caption{Photomodulation spectra of MeLPPP:PCBM blended films measured at RT.} \label{figure8}
\end{figure}

\begin{figure}
\unitlength1cm
\begin{picture}(4.0,8)
\put(-4.5,-0.5){ \epsfig{file=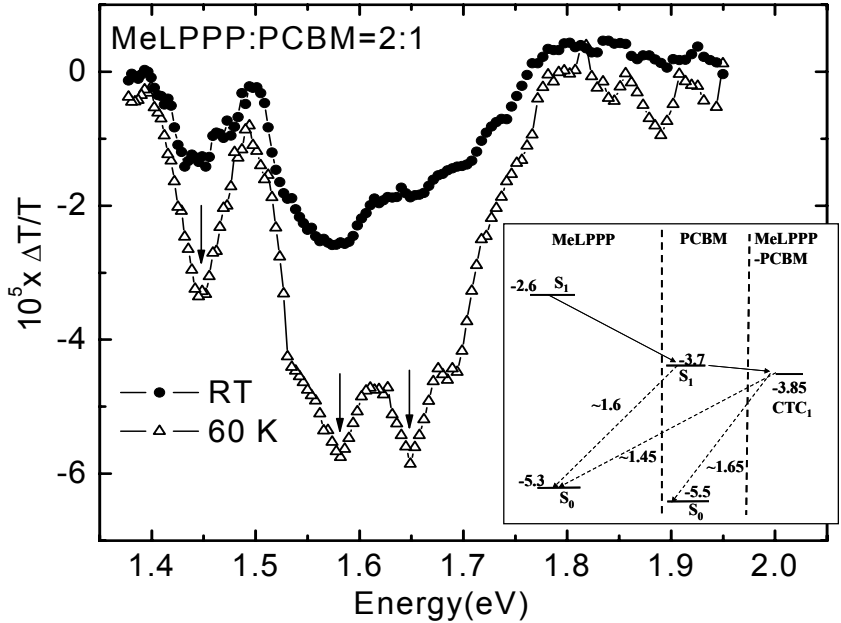, angle=0, width=12.cm, totalheight=8.7cm}}
\end{picture}
\caption{Photomodulation spectra of the 2:1 MeLPPPP:PCBM blended film at RT and 60 K. The inset shows the energy level diagram for the blend
with the dotted lines denoting the possible origin of emission peaks observed at 1.45 eV, 1.58, and 1.65 eV.} \label{figure9}
\end{figure}

The PIA spectra were measured from all three blend ratios used in the solar cells. The CTC emission in MeLPPP:PCBM is largely dependent upon the
PCBM concentration. The emission peaks at 1.45 eV and 1.6 eV are drastically reduced in the 1:1 and 1:2 blends, as seen in Fig. \ref{figure8}.
These measurements were all at RT. We further note that the 1.6 eV peak is better resolved at 60 K. There are at least two emission peaks at
1.58 eV and 1.65 eV along with the 1.45 eV emission, as shown in Fig. \ref{figure9}.

The inset of Fig. \ref{figure9} shows a plausible energy transfer mechanism in MeLPPP blended with PCBM. The triplet energy levels are not shown
here. The LUMO level energy offset in MeLPPP and PCBM may result in a direct energy transfer to the S$_1$ level of PCBM, followed by a charge
separation to the polaron pair. The other viable route is the formation of CTC states. As mentioned earlier, the S$_1$ to S$_0$ transition in
PCBM is symmetry forbidden. Most likely the emission features seen in the PIA spectrum result from the following transitions: 1.45 eV emission
is the transition from CTC$_1$ state to the ground state of MeLPPP, the 1.65 eV and 1.58 eV emission peaks are from the CTC$_1$ state to the
PCBM ground state and S$_1$ of PCBM to the ground state of MeLPPP, respectively.

The emission features in the PIA spectrum almost disappear for the 1:1 and 1:2 blended films. Increasing acceptor (PCBM) concentration should
enhance the dissociation rate of the bound polaron pair. Since the effective dielectric permittivity of the blend increases with PCBM
concentration, it reduces the energy barrier for CTCs enabling the formation of weakly-bound polaron pairs, \cite{smith} contributing to charges
in a solar cell. The PIA results of the blends correlate very well with the solar cell PCE. The 1:2 sample, which gives the highest PCE (Table
I), shows almost no radiative decay process; thus the separation of charges in this sample is most effective.

\subsection{PIA results of PhLPPP:PCBM}\label{sec:phlppp-pcbm film}
\begin{figure}
\unitlength1cm
\begin{picture}(4.0,7)
\put(-4.5,-1){ \epsfig{file=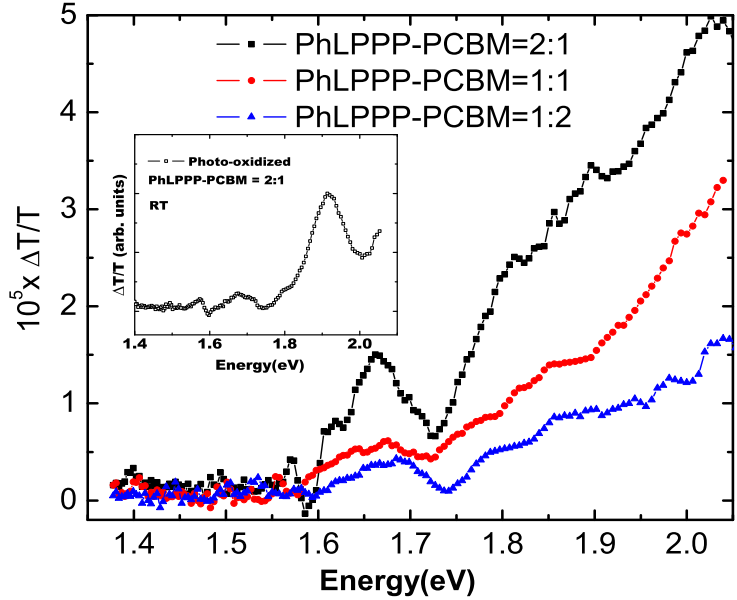, angle=0, width=11.cm, totalheight=9.7cm}}
\end{picture}
\caption{Photomodulation spectra of PhLPPP:PCBM blended films measured at RT. The inset is the photomodulation spectrum of a 2:1 photo-oxidized
PhLPPP:PCBM film. The 1.9 eV peak corresponds to the P2 polaron absorption.} \label{figure10}
\end{figure}

\begin{figure}
\unitlength1cm
\begin{picture}(4.0,8)
\put(-4.5,-1){ \epsfig{file=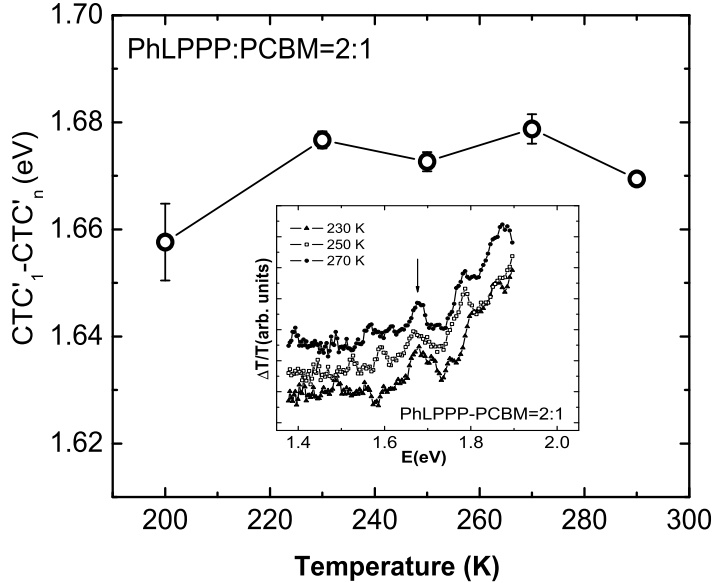, angle=0, width=11.cm, totalheight=9.7cm}}
\end{picture}
\caption{Temperature dependence of the CTC$^\prime_1$-CTC$^\prime_n$ absorption peak in the 2:1 PhLPPP:PCBM film. The inset shows a few
representative PIA spectra at selected temperatures for the same film.} \label{figure11}
\end{figure}

Figure \ref{figure10} shows the PIA spectra of PhLPPP:PCBM blended films measured at RT. The signal is positive in all cases. The main feature
here is an absorption peak at 1.65 eV, which systematically decreases with increasing PCBM concentration. This trend is also consistent with the
overall PCE in solar cells; the 1:1 and 1:2 samples show the highest efficiencies. Unfortunately, the polaron signature at 1.9 eV is very weak
at RT. It is however observable when the sample is photo-oxidized, as seen in the inset Fig. \ref{figure10} for the 2:1 sample. The polaron
absorption peak is seen in other blend ratios as well when photo-oxidized. Since the data are from photo-oxidized samples, it is hard to
quantify and compare the strength of the polaron signature for the various blend ratios as a direct correlation to the overall PCE.  Although
weak, the CTC state is still seen in the photo-oxidized samples, and the peak position of the CTC absorption is roughly the same as that for the
non-photo-oxidized samples. The CTC absorption in the 2:1 sample was further measured at different temperatures between 200-300 K; the inset of
Fig. \ref{figure11} shows a few representative spectra. The peak positions hardly change within this temperature range (Fig. \ref{figure11}),
indicating that the CTC state is localized. It has been recently shown that the diffusivity of nonthermalized triplet excitons in pristine
PhLPPP is temperature independent.\cite{reufer} The temperature independence of the CTC absorption in PhLPPP:PCBM blends therefore alludes to a
triplet origin to these charge transfer complexes.

\section{Discussion}\label{sec:discussion}
Our experimental results clearly show differences in the PIA spectra
of triplet-enhanced versus non-triplet-enhanced LPPP polymers
blended with PCBM. As pointed out the two polymers are structurally
identical with similar UV-vis PL and absorption properties.
Additionally, the morphology of the PhLPPP:PCBM and MeLPPP:PCBM
blends are similar as inferred from atomic force microscopy images
of the blended films. Both polymer blends show the formation of CTC
states but the nature and energy levels of these states are very
different in PhLPPP:PCBM and MeLPPP:PCBM. The formation of CTC
states in polymer/PCBM films depend on the ionization potential (IP)
of the polymer. Since the IP of PhLPPP and MeLPPP are identical, the
differences in CTC states must then originate from the presence of
triplet excitons in PhLPPP.

In MeLPPP:PCBM blends the emission peaks observed at 1.45 eV and 1.6 eV indicate that the CTC state must be slightly above the T$_1$ state of
PCBM (-3.85 eV). The CTC states that give rise to emission are typically singlet in character. \cite{cook,ichida} Hence energy transfer to the
CTC state occurs directly from the MeLPPP or PCBM singlet states. The dependence of the CTC emission on PCBM concentration clearly shows that
the origin of the CTC is from the blend itself and not from PCBM. Moreover, the photomodulation spectrum of a pure PCBM film was measured and no
features were observed in this energy range. The work by Veldman \emph{et al.} indicate the presence of nano-crystalline domains of PCBM; higher
PCBM concentration results in larger crystalline domains, enabling the dissociation of CTC states.\cite{veldman} The radiative emission
correlates very well with the solar cell performance; the MeLPPP blend ratio (1:2) that shows the highest PV efficiency shows the weakest
emission in PIA.

Since PhLPPP and MeLPPP have similar HOMO/LUMO energies, the CTC states arising from singlet excitons should have comparable energies. The PIA
spectroscopy from PhLPPP:PCBM clearly indicate an absorption peak at 1.65 eV, which is attributed to the difference in energy between
CTC$^\prime_1$ and a higher lying CTC$^\prime_n$ state indicated in Fig. \ref{figure7}. This may be due to long-lived CTC triplet states that
are directly populated from the triplet states of PhLPPP. We point out that the T$_1$$\longrightarrow$ T$_n$ transition energy in PCBM/C$_{60}$
is close to 1.6 eV.\cite{ichida,matsushita} It is difficult to distinguish between intramolecular CTC states in PCBM due to the presence of the
polymer and intermolecular CTC states in PhLPPP:PCBM blend. In a pump-probe setup an induced absorption peak at 1.65 eV in C$_{60}$ is observed
only in solution; the intra-molecular-type triplet exciton at 1.6 eV is not observed in a thin film of C$_{60}$.\cite{ichida} We do not observe
any feature at 1.65 eV from a pure  PCBM film in our PIA experiment. Since an absorption feature is seen rather than any radiative decay in PIA,
most likely the CTC$^\prime_1$ state lies below the singlet/triplet state of PCBM and the polymer for a favorable energy transfer process.

The TT absorption in the polymer (at 1.3 eV) was not observable in PhLPPP:PCBM blends at RT. A plausible explanation may be that the lifetimes
of the triplet excitons are reduced to the extent that a PIA experiment with a mechanical chopper is not adequate to observe the TT absorption
in the polymer. Additionally, the triplet states may be involved in the formation of the CTC triplet states in the blend, which also results in
these states being more localized. Had the triplet states of the polymer played no role in the formation of CTC states, one would expect similar
PIA spectra for PhLPPP and MeLPPP blended with PCBM. Furthermore, since PhLPPP:PCBM blends show a much higher PV efficiency compared to the
non-triplet-enhanced polymer, the CTC triplet states formed in these blends are most likely responsible for the weakly bound polaron-pairs,
contributing to charges.

\section{Conclusion}\label{sec:conclusion}
In conclusion, we observe the formation of CTC states in ladder-type PPP polymers blended with PCBM for solar cell applications via
photo-induced absorption studies. The nature of these CTC states is different for the triplet-enhanced PhLPPP compared to the
non-triplet-enhanced polymer, MeLPPP. The CTCs in PhLPPP:PCBM blend have a triplet character to it, which may be populated directly by the
lowest triplet state of the polymer. These states may be converted to weakly-bound polaron pairs, contributing to additional charges, and thus
enhancing the photovoltaic efficiency. In comparison, MeLPPP:PCBM blends show the presence of CTC states that are formed as a result of the
singlet excitons, which further lead to a radiative recombination. The CTC states are key to charge separation and efficiency of solar cells.
The possibility of triplet-enhanced power conversion efficiency in solution-processed LPPP-based polymers and PCBM organic solar cells will
stimulate future work on the development of other polymers with trace quantities of heavy metal and further investigation of CTC states for
improving photovoltaic efficiencies of polymer-based solar cells.


\begin{acknowledgments}
We gratefully acknowledge the support of this work through the National Science Foundation under grant No. ECCS-0823563.

\end{acknowledgments}


\end{document}